\magnification=1200 \baselineskip=13pt \hsize=16.5 true cm \vsize=21 true  cm
\def\parG{\vskip 10pt} \font\bbold=cmbx10 scaled\magstep2
\centerline{{\it Int. J. Mod. Phys.} {\bf C11}(7) (2000)}\parG
\centerline {\bbold Distribution of Traffic Penalties in Rio de Janeiro}\parG
\centerline{Rafaella A. N\'obrega, Cricia C. Rodegheri and Renato
C. Povoas}\parG
Instituto de F\'\i sica, Universidade Federal Fluminense\par
av. Litor\^anea s/n, Boa Viagem, Niter\'oi RJ, Brazil 24210-340\par
e-mail RAFAELLA@IF.UFF.BR\par

\vskip 0.3cm\leftskip=1cm\rightskip=1cm 

{\bf Abstract}\par

        Brazilian drivers caught in traffic violations accumulate points
in their official personal files. Here, we analyse the distribution
probability of these data, for the state of Rio de Janeiro, where 4199
drivers accumulated 20 or more penalty points during one year.\par

\leftskip=0pt\rightskip=0pt\parG


\vskip 15pt

        Some apparently complex systems have regularities that can be
fitted by long-tailed power-law probability distributions, instead of the
usual short-tailed exponential decay. A lot of examples are listed, for
instance, in [1]. In the present work, we searched to identify such
regularities in the distribution of penalties by infractor drivers.

        Brazilian traffic laws establish a grade which varies depending on
the infraction commmitted. Drivers who add 20 or more points during one
year are supposed to have their licenses suspended, loosing so their right
to drive during the next year. The (anonymous) list of these people for
the state of Rio de Janeiro was recently published in a widespread
newspaper [2]. It has 4199 entries, and the number of points for each
driver varies between 20 and 373.

        First, we have counted the number $n$ of drivers with a given
grade, i.e. a given number of points. These data are shown in Figure 1,
according to a double-log scale. Would this distribution fit to a
power-law, then the points would follow a straight line on this plot. As
one can see on this first plot, this is not so far from the truth, in
spite of the large data fluctuations observed. However, we can go further
analysing the same data more carefully.

        We have also counted the number $N$ of drivers with {\bf more
than} a certain number of points, i.e. the so-called accumulated
distribution. This quantity is related to the previous one as

$$N(p) = \sum_p^\infty n(x) \sim \int_p^\infty n(x) dx\,\,\,\, .$$

\noindent Would $n(x)$ follow a power law, i.e. $n(x) \sim x^{-\gamma}$,
then so would be the mathematical behaviour of $N(p)$, namely $N(p) \sim
p^{1-\gamma}$. However, the accumulated counting $N$ presents less
fluctuating data than $n$. Indeed, Figure 2 shows $N(p)$, with a much
better defined trend than Figure 1. Let's comment on two features of this
plot. First, the points on the right clearly do not follow a straight
line. However, {\bf each horizontal plateau corresponds to a single
driver}, with a no-drivers gap along the whole plateau. This region
corresponds to the {\bf very few} drivers with a large number of points,
namely a dozen among 4199, for which the statistics is not trustable.
Thus, let's disregard this minority, and analyse the left part of the plot
in Figure 2, where no plateau appears, for a number of points less than,
say, 80. The second feature of this plot is the small ``knee'' appearing
near 40 points, which invalidates the power-law behaviour. At most, we can
interpret the data as following {\bf two} distinct power-laws, with
different slopes (exponents) $1-\gamma \approx -3.5\,$ below 40 points,
and $1-\gamma' \approx -5.9\,$ above. However, we can go further,
performing a better yet analysis of the data.

        We can classify the drivers according to their ranks: number 1 is
the champion of infractions, with 373 points; number 2 has 158 points;
number 3 has 114; and so on. After the 11 higher infractors, we have some
ties: number 12 presents 71 points, as well as number 13 who accumulated
also 71 points. Numbers 18, 19, 20 and 21 share 65 points each. We do not
need to break these ties, but simply to arrange them according to any
order. In this way, the rank runs from 1 to 4199, exactly the total number
of drivers. Figure 3 shows the so-called Zipf's plot of our data, i.e. the
number of points as a function of the rank, for all 4199 infractor
drivers. This approach was introduced in the twenties by G.K Zipf [3], who
analysed the frequency of words appearing in books, American cities
classified according to their size, etc. Many other data were classified
in this way, as the most accessed internet sites, scientific citations,
goal makers in soccer championships, etc (for recent data, see, for
instance, [4,5]).

        Again because of the poor statistics, we can not trust in the data
relative to the few first ranked drivers, now positioned on the left part
of Figure 3. However, the curvature observed at the right part of the
figure clearly shows that we are not dealing with a power-law
distribution.

        Another option which was also recently observed in many systems is
the so-called stretched exponential (see, for instance, [6])

$$P = K \exp{(-b\, r^a)}\,\,\,\,\,\,   (a < 1)\,\,\,\, ,$$

\noindent which does not present infinitely-ranged tails as a power-law,
but also does not decay as fast as an exponential probability
distribution. It represents an intermediate behaviour. We have tried to
fit our data with this form, by fixing tentative values for the exponent
$a$, and plotting the number of points as a function of the quantity $r^a$
($r$ is the ranking) according to a linear-log scale. Figure 4 shows the
result for the best straight line fit we got, the continuous line, with $a
= 0.20$. We decide to skip out the first 20 ranked drivers from the fit,
the same few already quoted for which the statistics is not trustable.
Nevertheless, they also appear on the plot, clearly deviating from the
fitted line. The coefficient of correlation for this fit is also shown, as
well as the values for $K$ and $b$.

        In short, we have analysed the probability distribution for
traffic penalties in Rio de Janeiro state, Brazil. The data is well
represented by a stretched exponential, Figure 4, provided one rules out
the first higher infractor drivers ($\sim 20$ out of a total of 4199).

        We are indebted to Dietrich Stauffer who found the data in the
quoted Brazilian newspaper, and proposed us this exercise (we are
first-year students of undergraduate Physics course), and Paulo Murilo
Castro de Oliveira and Thadeu Josino Pereira Penna who helped us to
analyse the data.

\vskip 15pt
{\bf References}\parG

\item{[1]} P. Bak, {\sl How Nature Works: the Science of Self-Organized
Criticality}, Oxford University Press (1997).\par

\item{[2]} O Globo (Brazilian newspaper), Rio de Janeiro, October 13, page
11 (2000).\par

\item{[3]} G.K. Zipf, {\sl Human Behaviour and the Principle of Least
Effort}, Addison-Wesley, Cambridge, MA (1949)

\item{[4]} C. Tsallis and M.P. Albuquerque, {\it Eur. Phys. J.} {\bf B13},
777 (2000).\par

\item{[5]} L.C. Malacarne and R.S. Mendes, {\it Physica} {\bf A286}, 391
(2000).\par

\item{[6]} D. Sornette, {\sl Critical $\,$ Phenomena $\,$ in $\,$ Natural
$\,$ Sciences $\,$ (Chaos, Fractals, Self-organization and Disorder:
Concepts and Tools}, Springer Series in Synergetics, Heidelberg
(2000).\par

\vskip 15pt
\parG{\bf Figure Captions}\parG

\item{Figure 1} Crude data, counting the number of drivers with a given
number of penalty points.\par

\item{Figure 2} Accumulated distribution, counting the number of drivers
with more than a given number of points.\par

\item{Figure 3} Zipf's plot, where all 4199 drivers appear, ranked
according to their numbers of penalty points.\par

\item{Figure 4} Best fit of the data to a stretched exponential,
discarding the 20 higher infractor drivers.\par

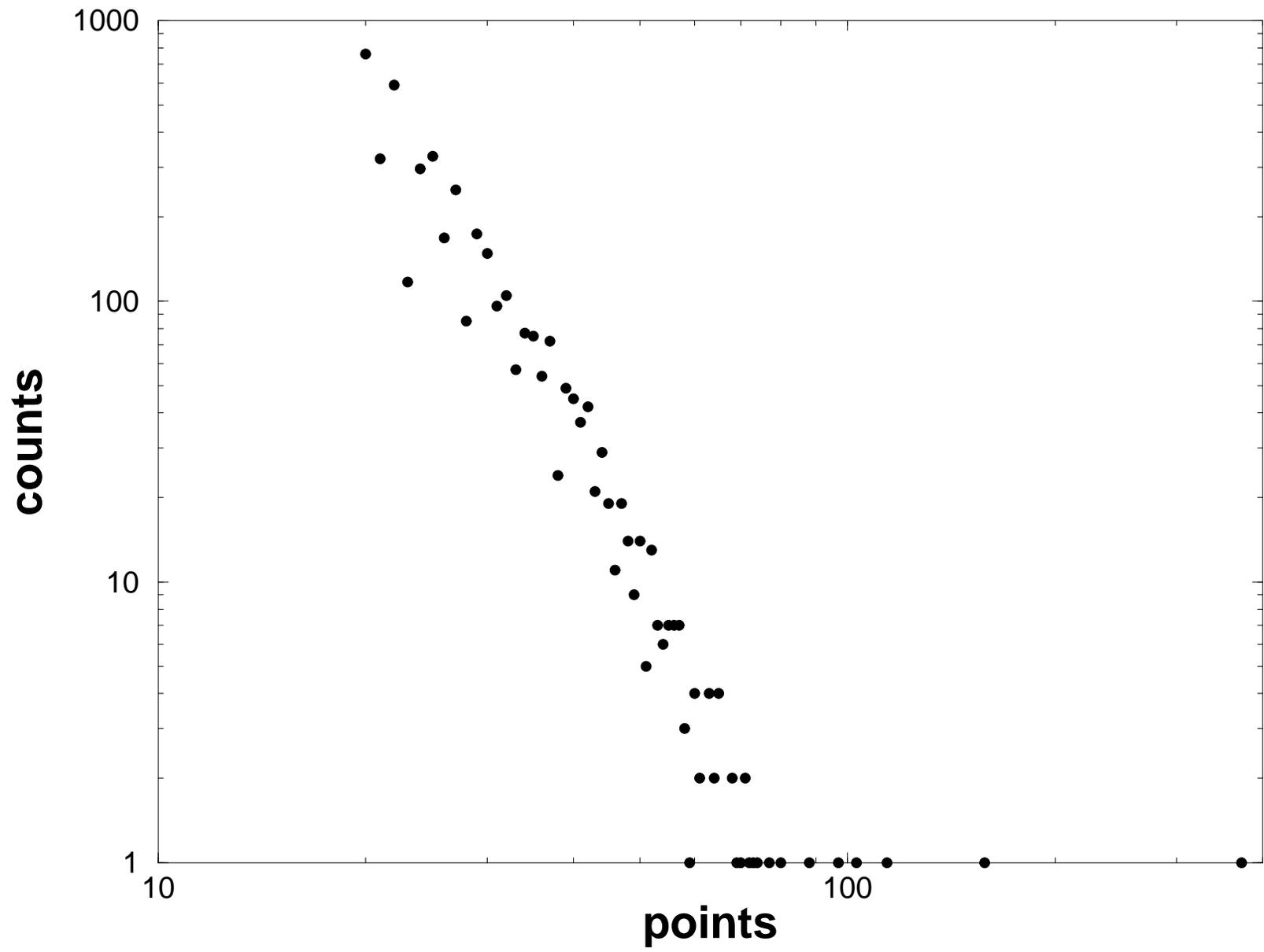
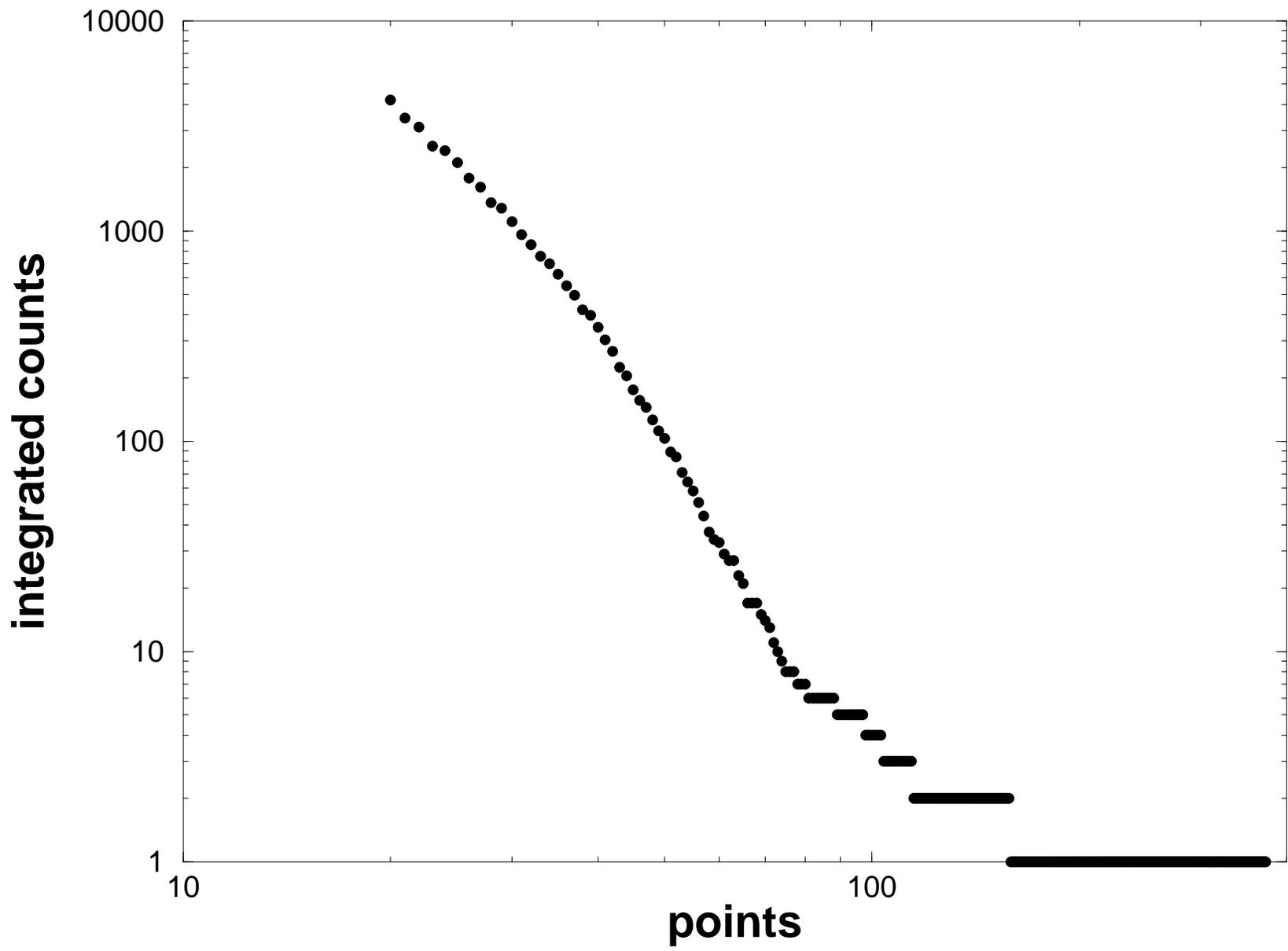
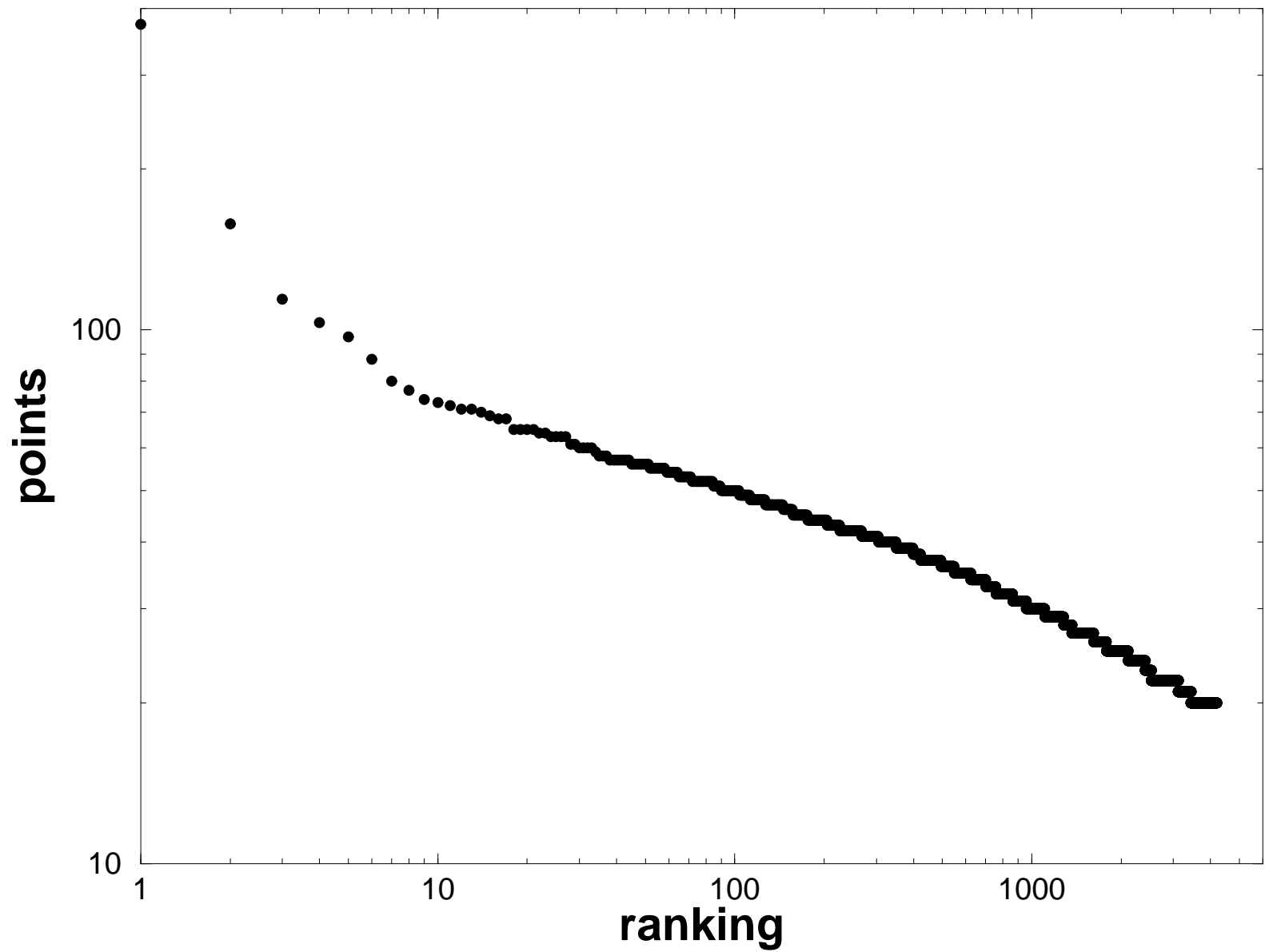
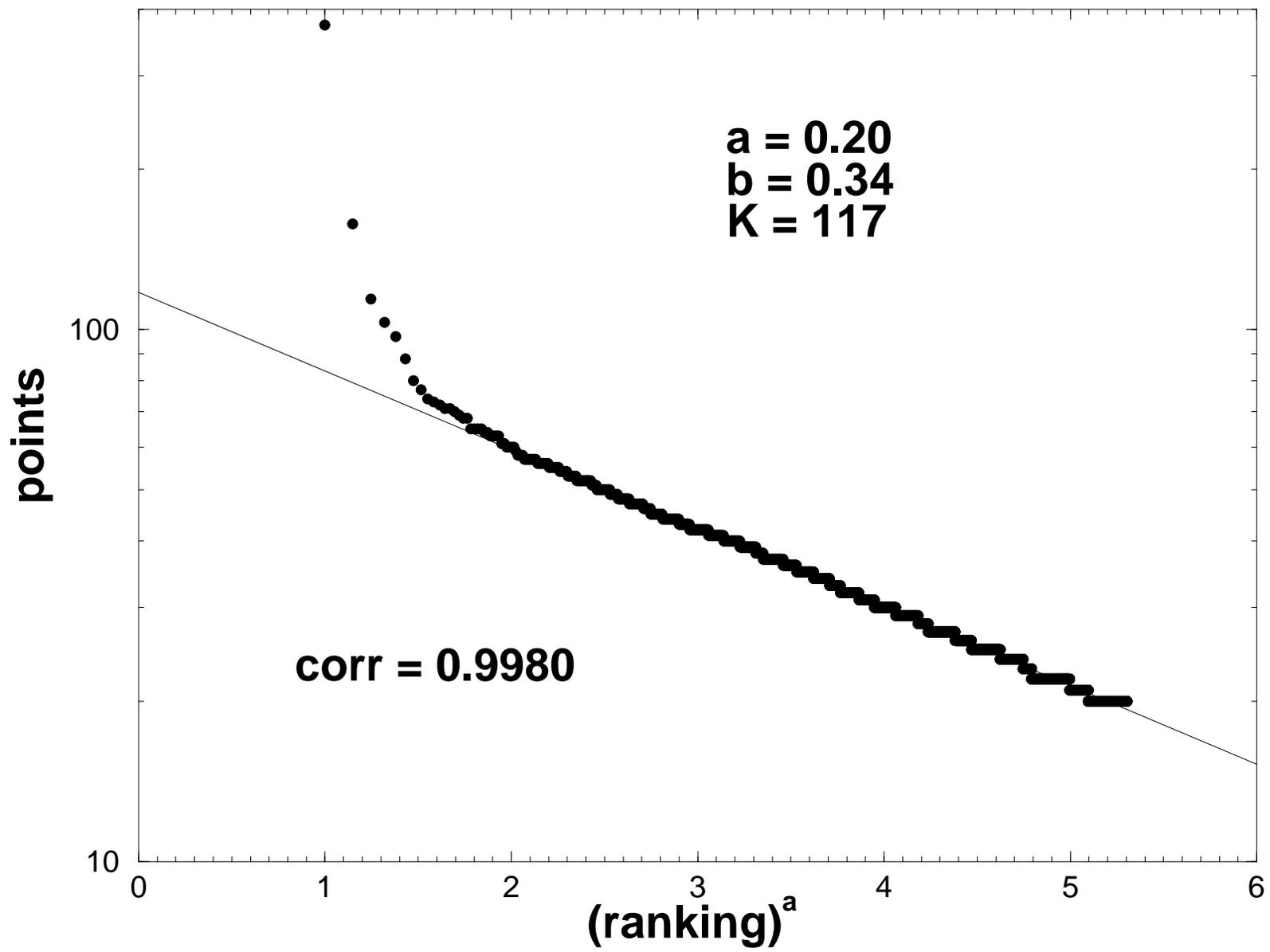
\bye